\documentclass[a4paper]{article}
\usepackage{amsmath}
\usepackage{latexsym}

\title{A Gauge field Induced by the Global Gauge Invariance of Action
Integral}

\author{Zaixing Huang\\[5pt]
  College of Aerospace Engineering\\
  Nanjing University of Aeronautics and Astronautics\\
  Nanjing 210016, P R China\\ E-mail: huangzx@nuaa.edu.cn}

\newcommand{\dLdya}{\frac{\partial{\mathit{L}}}{\partial{\mathit{\varphi}_{\alpha}}}}
\newcommand{\dLdyav}{\frac{\partial{\mathit{L}}}{\partial{\mathit{\varphi}_{\alpha,\nu}}}}

\begin{document}

\maketitle

\begin{abstract}
\noindent As a general rule, it is considered that the global gauge
invariance of an action integral does not cause the occurrence of
gauge field. However, in this paper we demonstrate that when the
so-called localized assumption is excluded, the gauge field will be
induced by the global gauge invariance of the action integral. An
example is given to support this conclusion.\\
\textbf{Key words:} gauge invariance, localized assumption, nonlocal
residual, gauge field
\end{abstract}

%  The paper
\section{Introduction}
Should it necessarily be the case that a statement holds for every
part of a body or field if it holds for the whole body or field?
The answer is in general negative. The examples in physics
provided by Edelen \cite{s1} show that an integral statement for
the whole body or field (For shorthand, we will use the "body" and
"field" without distinction in the latter) is true, and yet when
exactly the same statement is made for a subset of the body it
ceases to be valid. In fact, there are some physical phenomena in
which it is not always advisable to write a mathematical
representation for a part of a body that has the same form and
uses the same functions as occur in the corresponding formula for
the whole body. As a result, the assertion that a statement on a
body as a whole is valid to each part of the body is merely an
assumption in physics. ---This is the so-called localized
assumption \cite{s1,s2,s3}, which manifest itself in the following
procedures: \\(i) Statement of a global equilibrium for a state of
the body or field, that is

\begin{equation}\label{1}
\int\limits_\Omega
\underbar{\itshape{T}}(\underbar{\itshape{x}})\mathrm{d}V(\underbar{\itshape{x}})=0.      %(1)
\end{equation}

where {\underbar{\itshape{x}}} denotes the space-time coordinate.
{\underbar{\itshape{T}}}({\underbar{\itshape{x}}}) is a state
function, which can be either scalar, vector or tensor depending on
circumstance. $\Omega$ is a bounded space-time domain occupied by
the body or field.\\
(ii) Assume that Eq.(\ref{1}) is also valid for every part
{\itshape{V}} of with the same function
{\underbar{\itshape{T}}}(\underbar{\itshape{x}}), that is

\begin{equation}\label{2}
\int\limits_V
\underbar{\itshape{T}}(\underbar{\itshape{x}})\mathrm{d}V(\underbar{\itshape{x}})=0.       %(2)
\end{equation}

(iii) In terms of the localized theorem \cite{s4}, the local
equation can be, from Eq.(\ref{2}), given as follows:
\begin{equation}\label{3}
\underbar{\itshape{T}}(\underbar{\itshape{x}})=0.                 %(3)
\end{equation}
The step from (i) to (ii) is just an embodiment of the localized
assumption. This assumption has been adopted all along in physics.
By means of it, an integral representation can be conveniently
transformed into a differential equation. If the localized
assumption is abandoned, then a way will lead to the so-called
nonlocal theories in which the relevant physical formulations are
generally given by a group of
integro-differential equations.\\
At present, there are two ways to comprehend the invariance of the
Lagrange field. One is based on the Lagrangian; the other is based
on the action integral of the Lagrangian \cite{s5}. For the gauge
transformation, only the invariance of the Lagrangian is concerned
in literatures. Maybe it is due to the fact that under the global
gauge transformation, the invariance of the Lagrangian is regarded
to be equivalent to the invariance of the action integral.
However, if an elaborate analysis is made, one will find that the
equivalence between the invariance of the Lagrangian and that of
the action integral is guaranteed by the localized assumption.
According to this assumption, the invariance of the action
integral defined on a domain as a whole also inevitably holds for
any part of this domain no matter how small it is. Therefore, some
problems to be worth asking are what is the reason for needing
such an assumption, and what will happen when the localization
hypothesis is excluded. ---To answer these questions is the
subject of this
paper.\\
The premise of this paper contains three main propositions: 1) A
body or physical field is supposed to distribute over a bounded
space-time domain; 2) Under the gauge transformations, the
invariance of a Lagrangian system should be comprehended as the
invariance of the action integral of the Lagrangian, not the
Lagrangian; 3) The localized assumption is considered to be no
avail. On the basis of these premises, emphasis of this paper is
focused on how to express the gauge field induced by the global
gauge invariance of the action integral after the localized
assumption fails, and the connection between the gauge field and
the conservation flux.\\
The paper is divided into five parts. The first section is an
introduction, which gives the background of this paper. In the
second section, we discussed that under the condition of abandoning
the localized assumption, the connection between the invariance of
action integral and the conservation flux. The nonlocal balance
equation of the conservation flux is established by introducing the
nonlocal residual. In the third section, the local gauge invariance
of action integral is studied under the local gauge transformation,
the gauge field is introduced, and then it is extended to the case
of the global gauge transformation. ---On the basis of this, the
correlation between the nonlocal residual and the gauge field is
determined. In the fourth section, a complex scalar field as an
example is used to show the limitation of the localized assumption.
By means of the nonlocal balance equation of the complex scalar
field, an explicitly relation between the nonlocal residual and the
gauge field is given. Finally, some discussions on the results
obtained in this paper are drawn.

\section{Global gauge invariance of the action integral}
Suppose that ${\mathit{x}^{\mu}}$  ($\mu$ =1, 2, бн,
{\itshape{k}}) and ${\mathit{\varphi}_{\alpha}}$ ($\alpha$=1, 2,
бн, {\itshape{n}}) are the space-time coordinates and the
variables of field, respectively. The action integral
$\mathit{A}[\mathit{\varphi}_{\alpha}] $, defined on a bounded
space-time domain $\Omega \subset \mathit{E}^n$, takes the form as
follows

\begin{equation}\label{4}
\mathit{A}[\mathit{\varphi}_{\alpha}]=\int\limits_\Omega
\mathit{L}(\mathit{x}^{\mu}, \mathit{\varphi}_{\alpha},
\mathit{\varphi}_{\alpha,\nu})\mathrm{d}V(\mathit{x}^{\mu}),
\end{equation}          %(4)

where $\mathit{L}(\mathit{x}^{\mu}, \mathit{\varphi}_{\alpha},
\mathit{\varphi}_{\alpha,\sigma})$ is the Lagrangian density
function, or simply called the Lagrangian. Consider an
infinitesimal gauge transformation
\begin{equation}\label{5}
\mathit{\varphi}_{\alpha}(\mathit{x}^{\mu}) \to
\mathit{\tilde\varphi}_{\alpha}(\mathit{x}^{\mu})=\mathit{\varphi}_{\alpha}(\mathit{x}^{\mu})+
\mathit{\delta}\mathit{\varphi}_{\alpha}(\mathit{x}^{\mu}).
\end{equation}                                               %(5)
The action integral $\mathit{A}[\mathit{\varphi}_{\alpha}] $is
said to be gauge symmetry if it is form-invariant with respect to
the infinitesimal gauge transformation, i.e.,

\begin{equation}\label{6}
\int\limits_\Omega \mathit{L}(\mathit{x}^{\mu},
\mathit{\tilde{\varphi}}_{\alpha},
\mathit{\tilde{\varphi}}_{\alpha,\nu})\mathrm{d}V(\mathit{x}^{\mu})=\int\limits_\Omega
\mathit{L}(\mathit{x}^{\mu}, \mathit{\varphi}_{\alpha},
\mathit{\varphi}_{\alpha,\nu})\mathrm{d}V(\mathit{x}^{\mu}).
\end{equation}                                     %(6)

After rearrangement, Eq.(\ref{6}) becomes

\begin{equation}\label{7}
\int\limits_\Omega [\mathit{L}(\mathit{x}^{\mu},
\mathit{\tilde{\varphi}}_{\alpha},
\mathit{\tilde{\varphi}}_{\alpha,\nu})-
\mathit{L}(\mathit{x}^{\mu}, \mathit{\varphi}_{\alpha},
\mathit{\varphi}_{\alpha,\nu})]\mathrm{d}V(\mathit{x}^{\mu})=0.
\end{equation}                                     %(7)

It is easy to calculate that

\begin{gather}\label{8}
\begin{split}
 \delta \mathit{L}&=
\mathit{L}(\mathit{x}^{\mu},\mathit{\tilde{\varphi}}_{\alpha},\mathit{\tilde{\varphi}}_{\alpha,\nu})-
\mathit{L}(\mathit{x}^{\mu}, \mathit{\varphi}_{\alpha},\mathit{\varphi}_{\alpha,\nu})\\
&=\dLdya\mathit{\delta}\mathit{\varphi}_{\alpha}+\dLdyav\mathit{\delta}\mathit{\varphi}_{\alpha,\nu}\\
&=\dLdya\mathit{\delta}\mathit{\varphi}_{\alpha}+(\dLdyav\mathit{\delta}\mathit{\varphi}_{\alpha})_{,\nu}
-(\dLdyav)_{,\nu}\mathit{\delta}\mathit{\varphi}_{\alpha}\\
&=(\dLdyav\mathit{\delta}\mathit{\varphi}_{\alpha})_{,\nu}
+[(\dLdya-(\dLdyav)_{,\nu}]\mathit{\delta}\mathit{\varphi}_{\alpha}.
\end{split}
\end{gather}                                                        %(8)

Here, repeated indices mean summation. Substituting Eq.(\ref{8})
into (\ref{7}) yields

\begin{equation}\label{9}
\int\limits_\Omega
\{(\dLdyav\mathit{\delta}\mathit{\varphi}_{\alpha})_{,\nu}
+[(\dLdya-(\dLdyav)_{,\nu}]\mathit{\delta}\mathit{\varphi}_{\alpha}\}\mathrm{d}V(\mathit{x}^{\mu})=0.
\end{equation}                                           %(9)

Assume that the action integral
$\mathit{A}[\mathit{\varphi}_{\alpha}] $ takes an extremum on
${\mathit{\varphi}_{\alpha}}$. Then, the Lagrangian necessarily
satisfies the Euler-Lagrange equation (motion equation) below:

\begin{equation}\label{10}
\dLdya-(\dLdyav)_{,\nu}=0.              %(10)
\end{equation}

Inserting Eq.(\ref{10}) in (\ref{9}) leads to

\begin{equation}\label{11}
\int\limits_\Omega
(\dLdyav\mathit{\delta}\mathit{\varphi}_{\alpha})_{,\nu}
\mathrm{d}V(\mathit{x}^{\mu})=0.
\end{equation}                            %(11)

If Eq.(\ref{5}) belongs to a finite Lie group of infinitesimal
transformations, according to the representation of Lie group,
$\delta \varphi_\alpha$ can be written as \cite{s6,s7,s8}

\begin{equation}\label{12}
\mathit{\delta \varphi}_{\alpha} =\mathit{{\varepsilon}^{\beta}
{\Phi}_{\beta \alpha}},
\end{equation}                    %(12)

where $\mathit{{\varepsilon}^{\beta}}$ is an infinitesimal
parameter independent of the space-time coordinates and
$\mathit{{\Phi}_{\beta \alpha}}$ is the infinitesimal generator of
Lie group of transformations. Substituting Eq.(\ref{12}) into
(\ref{11}) yields

\begin{equation}\label{13}
\mathit{{\varepsilon}^{\beta}}\int\limits_\Omega
(\dLdyav\mathit{{\Phi}_{\beta \alpha}})_{,\nu}
\mathrm{d}V(\mathit{x}^{\mu})=0.
\end{equation}                        %(13)

Due to $\mathit{{\varepsilon}^{\beta}}$ taking an arbitrary value,
Eq.(\ref{13}) reduces to

\begin{equation}\label{14}
\int\limits_\Omega (\dLdyav\mathit{{\Phi}_{\beta \alpha}})_{,\nu}
\mathrm{d}V(\mathit{x}^{\mu})=0.
\end{equation}                           %(14)

If the localized assumption is true, then Eq.(\ref{14}) is also
valid for any ${\mathit{V} \subset \Omega}$. That is

\begin{equation}\label{15}
\int\limits_V (\dLdyav\mathit{{\Phi}_{\beta \alpha}})_{,\nu}
\mathrm{d}V(\mathit{x}^{\mu})=0.
\end{equation}                          %(15)

Thus, applying the localization theorem \cite{s4} to Eq.(\ref{15})
gives

\begin{equation}\label{16}
(\dLdyav\mathit{{\Phi}_{\beta \alpha}})_{,\nu} = 0.
\end{equation}                            %(16)

This is a result of the well-known Noether's theorem\cite{s6,s7}.
However, we have not yet a sufficient reason to need such a prior
condition as the localized assumption. Therefore, if the
localization assumption is no longer considered to be valid, then
we can not directly derive Eq.(\ref{16}) from (\ref{15}), instead,
a new term will occur in Eq.(\ref{15}) such that

\begin{equation}\label{17}
\int\limits_V (\dLdyav\mathit{{\Phi}_{\beta \alpha}})_{,\nu}
\mathrm{d}V(\mathit{x}^{\mu})=\mathit{R_\beta (V)}.
\end{equation}                                         %(17)

Obviously, $\mathit{R_\beta (V)}$ is a generalized measure function
defined on {\itshape{V}}. Assume it is absolutely continuous with
respect to V. So according to the Radon-Nikodym theorem \cite{s9},
$\mathit{R_\beta (V)}$ can be represented as

\begin{equation}\label{18}
\mathit{R_\beta (V)}= \int\limits_V \mathit{{F}_{\beta
}}(\mathit{x}^{\mu})\mathrm{d}V(\mathit{x}^{\mu}),
\end{equation}                                        %(18)

where $\mathit{{F}_{\beta }}=\mathit{{F}_{\beta
}}(\mathit{x}^{\mu})$ is called the nonlocal residual or
localization residual. Substituting Eq.(\ref{18}) into (\ref{17})
yields

\begin{equation}\label{19}
\int\limits_V (\dLdyav\mathit{{\Phi}_{\beta \alpha}})_{,\nu}
\mathrm{d}V(\mathit{x}^{\mu})=\int\limits_V \mathit{{F}_{\beta
}}(\mathit{x}^{\mu})\mathrm{d}V(\mathit{x}^{\mu}).
\end{equation}                                       %(19)

Since {\itshape{V}} can be arbitrarily chosen, so Eq.(\ref{19})
holds if and only if

\begin{equation}\label{20}
(\dLdyav\mathit{{\Phi}_{\beta \alpha}})_{,\nu} =\mathit{{F}_{\beta
}}.
\end{equation}             %(20)

Eq.(\ref{20}) is referred to as the nonlocal balance equation,
which is a generalization of the Noether's formulation under the
global gauge transformation. Let

\begin{equation}\label{21}
\mathit{{J}_{\beta }^{\nu}}=\dLdyav\mathit{{\Phi}_{\beta \alpha}},
\end{equation}

where $\mathit{{J}_{\beta }^{\nu}}$ is called the conservation
flux. Accordingly, Eq.(\ref{20}) can be also shortly written as

\begin{equation}\label{22}
\mathit{{J}_{\beta,\nu }^{\nu}}=\mathit{{F}_{\beta }}.
\end{equation}

When $\mathit{V=\Omega}$, comparison of Eq.(\ref{19}) with
(\ref{14}) leads to the so-called "zero mean condition",

\begin{equation}\label{23}
\int\limits_\Omega \mathit{{F}_{\beta
}}(\mathit{x}^{\mu})\mathrm{d}V(\mathit{x}^{\mu})=0.
\end{equation}

This equation shows that, although $\mathit{{F}_{\beta }}$ has
influences on the local conservation flux, its global effects on
$\Omega$ as a whole are null. Because $\mathit{{F}_{\beta }}$ does
not vanish everywhere, Eq.(\ref{23}) forms a constraint to
Eq.(\ref{20}) or (\ref{22}).

\section{Correlation between the nonlocal residual and the gauge field}
In physics, the nonlocal residual is considered to originate from
self-interactions among different local regions within a body or
field \cite{s1,s2}. These self-interactions induce, in the
sub-domain of $\Omega$ , the symmetry breaking of the action
integral. Therefore, the nonlocal residual can be regarded as a new
source of the conservation flux. On the other hand, the global gauge
symmetry of an action integral is not extended to the local symmetry
unless a gauge field is introduced by means of the Yang-Mills
minimal coupling principle \cite{s6,s7}. The new gauge field acts
also as a source of the conservation flux in the local gauge
invariance. Such facts hint us that there are probably some
correlations between the
nonlocal residual and the gauge field.\\
The infinitesimal gauge transformation can also be represented as
\cite{s6}:

\begin{equation}\label{24}
\mathit{\varphi}_{\alpha}(\mathit{x}^{\mu}) \to
\mathit{\tilde\varphi}_{\alpha}(\mathit{x}^{\mu})=(1+\mathit{\varepsilon^\beta
\Phi_\beta})\mathit{\varphi}_{\alpha}(\mathit{x}^{\mu}),
\end{equation}

where $\mathit{\varepsilon^\beta}$ is a linear operator, which is
written as

\begin{equation}\label{25}
\mathit{\Phi_\beta=\frac{\partial}{\partial{\varepsilon^\beta}}\bigg|_{{\varepsilon^\beta}=0}}.
\end{equation}

If $\varepsilon^\beta$ is independent of the space-time
coordinates, the transformation (\ref{24}) is called the global
gauge transformation. Or else, it refers to the local gauge
transformation. Under the global gauge transformation, the
derivative of the field variable $\mathit{\varphi_\alpha}$ with
respect to the coordinate $\mathit{x^\nu}$ has the same form as
Eq.(\ref{24}), i.e.,

\begin{equation}\label{26}
\mathit{\varphi}_{\alpha,\nu}(\mathit{x}^{\mu}) \to
\mathit{\tilde\varphi}_{\alpha,\nu}(\mathit{x}^{\mu})=(1+\mathit{\varepsilon^\beta
\Phi_\beta})\mathit{\varphi}_{\alpha,\nu}(\mathit{x}^{\mu}).
\end{equation}                                                  %(26)

If the action integral is unchanged with the global gauge
transformation \footnote[1]{If we do not take the localized
assumption into account, the invariance of action integral is not
equivalent to the invariance of Lagrangian under a global gauge
transformation.}, then Eq.(\ref{22}) will be given once again, and
the conservation flux can be written as

\begin{equation}\label{27}
\mathit{J_\beta^\nu}=\dLdyav\mathit{\Phi_\beta\varphi_\alpha}.
\end{equation}                                                   %(27)

If the localized assumption is supposed to be valid, Eq.(\ref{22})
will then reduce to

\begin{equation}\label{28}
\mathit{{J}_{\beta,\nu }^{\nu}}=0.
\end{equation}                                                    %(28)

However, under the local gauge transformation, neither
Eq.(\ref{22}) nor (\ref{28}) holds because the action integral is
no longer invariant, as

\begin{equation}\label{29}
\mathit{\varphi}_{\alpha,\nu}(\mathit{x}^{\mu}) \to
\mathit{\tilde\varphi}_{\alpha,\nu}(\mathit{x}^{\mu})=(1+\mathit{\varepsilon^\beta
\Phi_\beta})\mathit{\varphi}_{\alpha,\nu}(\mathit{x}^{\mu})+(\mathit{\varepsilon^\beta
\Phi_\beta})_{,\nu}\mathit{\varphi}_{\alpha}(\mathit{x}^{\mu}).
\end{equation}

To construct a local gauge invariant theory, a new field, called the
gauge field, should be introduced to render the action integral
invariant. According to the Yang-Mills minimal coupling principle
\cite{s7}, we define a covariant derivative as follows:

\begin{equation}\label{30}
\mathit{\frac{D}{Dx^\nu}=\frac{\partial}{\partial{x^\nu}}+eA_\nu},
\end{equation}

where {\itshape{e}} denotes the coupling constant. $A_\nu$ refers
to the gauge field, which transforms according to

\begin{equation}\label{31}
\mathit{A_\nu(x^\mu)\to
{\tilde{A}}_\nu(x^\mu)=A_\nu(x^\mu)+\varepsilon^{\beta}[\Phi_\beta,A_\nu]-\frac{\mathrm{1}}{e}\frac{\partial{(\varepsilon^\beta\Phi_\beta)}}{\partial{x^\nu}}},
\end{equation}

in which $[\Phi_\beta,A_\nu]$ is defined as

\begin{equation}\label{32}
\mathit{[\Phi_\beta,A_\nu]=\Phi_\beta A_\nu -A_\nu \Phi_\beta}.
\end{equation}

Under the local gauge transformation, it is easy to verify that

\begin{equation}\label{33}
\mathit{\frac{D \varphi_\alpha}{D x^\nu}\to
\frac{\tilde{D}\tilde{\varphi}_\alpha}{\tilde{D}
x^\nu}=(\mathrm{1}+\varepsilon^\beta \Phi_\beta)\frac{D
\varphi_\alpha}{D x^\nu}},
\end{equation}

which has the same form as Eq.(\ref{26}).\\
In terms of the Yang-Mills minimal replacing principle \cite{s7}, we
use the covariant derivative $D/{Dx^\nu}$ instead of the common
derivative ${\partial}/{\partial{x^\nu}}$ in the Lagrangian. As
thus, under the local gauge transformation, the action integral will
remain unchanged. That is,

\begin{equation}\label{34}
\int\limits_\Omega \mathit{L(x^\mu, \tilde{\varphi}_\alpha ,
\frac{\tilde{D}\tilde{\varphi}_\alpha}{\tilde{D}
x^\nu})}\mathrm{d}V(\mathit{x}^{\mu})=\int\limits_\Omega \mathit
{L ( x^ \mu,  \varphi _ \alpha, \frac{D \varphi_\alpha}{D
x^\nu}})\mathrm{d}V(\mathit{x}^{\mu}).
\end{equation}

In order to derive the conservation flux from Eq.(\ref{34}), we
need to calculate
\begin{align}\label{35}
&\mathit{\delta
L=L(x^\mu,\tilde{\varphi}_\alpha,\frac{\tilde{D}\tilde{\varphi}_\alpha}{\tilde{D}
x^\nu})- L(x^\mu, \varphi_\alpha,\frac{D \varphi_\alpha}{D x^\nu})} \notag\\
&=\mathit{\dLdya\delta\varphi_\alpha+\frac{\partial{L}}{\partial{(\dfrac{D
\varphi_\alpha}{D x^\nu}})}\delta (\frac{D \varphi_\alpha}{D x^\nu})}\notag\\
& =\mathit{\dLdya (\varepsilon^\beta \Phi_\beta \varphi_\alpha)+\dLdyav(\varepsilon^\beta\Phi_\beta
\frac{D \varphi_\alpha}{D x^\nu})}\notag\\
& =\mathit{\dLdya \varepsilon^\beta \Phi_\beta
\varphi_\alpha+\dLdyav\varepsilon^\beta\Phi_\beta
(\varphi_{\alpha,\nu}+eA_\nu \varphi_\alpha)}\notag\\
& =\mathit{\varepsilon^\beta\dLdya \Phi_\beta
\varphi_\alpha+\varepsilon^\beta\dLdyav\Phi_\beta
\varphi_{\alpha,\nu}+e \varepsilon^\beta\dLdyav\Phi_\beta A_\nu
\varphi_\alpha}\notag\\
& =\mathit{\varepsilon^\beta(\dLdyav)_{,\nu} \Phi_\beta
\varphi_\alpha+\varepsilon^\beta\dLdyav(\Phi_\beta
\varphi_{\alpha})_{,\nu}+e \varepsilon^\beta\dLdyav\Phi_\beta A_\nu
\varphi_\alpha}\notag\\
& =\mathit{\varepsilon^\beta(\dLdyav\Phi_\beta
\varphi_{\alpha})_{,\nu}+e
\varepsilon^\beta\dLdyav\Phi_\beta A_\nu \varphi_\alpha}\notag\\
& =\mathit{\varepsilon^\beta J^\nu_{\beta,\nu}+e
\varepsilon^\beta\dLdyav\Phi_\beta A_\nu \varphi_\alpha}.
\end{align}

The last equals sign is due to Eq.(\ref{27}). Inserting
Eq.(\ref{35}) in (\ref{34}) leads to

\begin{equation}\label{36}
\mathit{\int\limits_\Omega \varepsilon^\beta (J^\nu_{\beta,\nu}+e
\dLdyav\Phi_\beta A_\nu
\varphi_\alpha)\mathrm{d}V(\mathit{x}^{\mu})}=0.
\end{equation}

Because $\varepsilon^\beta$ in Eq.(\ref{36}) can be arbitrarily
choose, we have

\begin{equation}\label{37}
\mathit{J^\nu_{\beta,\nu}+e \dLdyav\Phi_\beta A_\nu
\varphi_\alpha}=0.
\end{equation}                                       %(37)

It is interesting to notice the distinguish of $\varepsilon^\beta$
in Eq.(\ref{13}) and in (\ref{36}). In Eq.(\ref{13}),
$\varepsilon^\beta$ is independent of the space-time coordinates.
So it can be moved into the exterior of integral symbol. ---This
makes us to derive Eq.(\ref{16}) from Eq.(\ref{14}) only by way of
the localization theorem \cite{s4}. On the contrary,
$\varepsilon^\beta$ in Eq.(\ref{36}) can not be moved into the
exterior of integral symbol due to it depending on coordinates.
Consequently, we can directly obtain Eq.(\ref{37}) from
Eq.(\ref{36}) in terms of the variational lemma \cite{s8}, not
needing to rely on the localization
theorem.\\
In fact, Eq.(\ref{37}) also holds for the global gauge
transformation. Under this circumstance, $\varepsilon^\beta$ is a
constant, and Eq.(\ref{31}) reduces to

\begin{equation}\label{38}
\mathit{A_\nu(x^\mu)\to
{\tilde{A}}_\nu(x^\mu)=A_\nu(x^\mu)+\varepsilon^{\beta}[\Phi_\beta,A_\nu]}.
\end{equation}                                      %(38)

When $\varepsilon^\beta$ is an infinitesimal constant, the second
term of Eq.(\ref{37}) also satisfies the zero mean condition. In
order to verify this argument, we firstly need to prove that the
integral of $\mathit{J^\nu_{\beta,\nu}}$in Eq.(\ref{37}) on
$\Omega$ is equal to zero. For this, let us write out the
necessary condition of the action integral taking the extrema,

\begin{equation}\label{39}
\mathit{\int\limits_\Omega
[\dLdya-(\dLdyav)_{,\nu}]\delta\varphi_\alpha\mathrm{d}V(x^{\mu})+\int\limits_{\partial{\Omega}}
\dLdyav\delta\varphi_{\alpha} n_\nu \mathrm{d}S(x^{\mu})}=0.
\end{equation}                       %(39)

On the boundary of $\Omega$, regardless of whether
$\delta\varphi_{\alpha}$ is zero or not zero, Eq.(\ref{10}) is
always valid. Accordingly, Eq.(\ref{39}) is simplified to

\begin{equation}\label{40}
\mathit{\int\limits_{\partial{\Omega}}
\dLdyav\delta\varphi_{\alpha} n_\nu \mathrm{d}S(x^{\mu})}=0.
\end{equation}                       %(40)

Applying the divergence theorem to Eq.(\ref{40}) leads to

\begin{equation}\label{41}
\mathit{\int\limits_\Omega (\dLdyav\delta\varphi_{\alpha})_{,\nu}
\mathrm{d}V(x^{\mu})}=0.
\end{equation}                       %(41)

Because $\delta\varphi_\alpha$ is arbitrary, a selection is let
$\mathit{\delta\varphi_\alpha=\varepsilon^\beta \Phi_\beta
\varphi_\alpha}$ and let $\varepsilon^\beta$ be an infinitesimal
constant. As a result, Eq.(\ref{41}) reduces to

\begin{equation}\label{42}
\mathit{\int\limits_\Omega (\dLdyav\Phi_\beta
\varphi_{\alpha})_{,\nu} \mathrm{d}V(x^{\mu})}=0.
\end{equation}                       %(42)

By virtue of Eq.(\ref{27}), Eq.(\ref{42}) can be also written as

\begin{equation}\label{43}
\mathit{\int\limits_\Omega J_{\beta ,\nu}^\nu
\mathrm{d}V(x^{\mu})}=0.
\end{equation}                       %(43)

Taking the integral for Eq.(\ref{37}) on $\Omega$ and using
Eq.(\ref{43}), we have

\begin{equation}\label{44}
\mathit{\int\limits_\Omega e \dLdyav\Phi_\beta A_\nu
\varphi_\alpha\mathrm{d}V(x^{\mu})}=0.
\end{equation}                                  %(44)

This shows that the integrand in Eq.(\ref{44}) also satisfies the
zero mean condition. ---By way of this conclusion, comparing
Eq.(\ref{37}) with (\ref{22}) will lead to

\begin{equation}\label{45}
\mathit{F_\beta= e \dLdyav\Phi_\beta A_\nu \varphi_\alpha}.
\end{equation}                                  %(45)

Therefore, the nonlocal residual surely has a natural connection
with the gauge field. In general, when the localized assumption is
available, it is meaningless in physics to introduce the gauge
field to describe the global gauge invariance. However, if the
localized assumption fails, then the gauge field induced by the
global gauge invariance physically becomes feasible. It can be
used to characterize the nonlocal residual, just as seen from
Eq.(\ref{45}).

\section{An example: A gauge induced by the global gauge invariance of action integral }
For convenience, in this section we will use the following
notations:

\begin{equation}\label{46}
\partial_\mu=\frac{\partial}{\partial
x^\mu},\mspace{16mu} \partial^\mu=g^{\mu\nu}\partial_\nu ,
\end{equation}                           %(46)
\begin{equation}\label{47}
\Box=\partial_\mu
\partial^\mu=\frac{1}{c^2}\frac{\partial^2}{\partial t^2}-\nabla^2 ,
\end{equation}                            %(47)

where $g^{\mu\nu}$ refers to the metric tensor and $\Box$ denotes
the d' Alembertian operator.\\
Consider a complex scalar field. Because the action integral should
be real, so the Lagrangian of this complex scalar field is supposed
to have the form below:

\begin{equation}\label{48}
L=\partial_\mu \varphi \partial^\mu
\varphi^*-m^2\varphi\varphi^*-V(\varphi\varphi^*)-U(\partial_\mu
\varphi \partial^\mu \varphi^*).
\end{equation}                        %(48)

Here, $\varphi$  and $\varphi^*$ are a pair of conjugate complex
variables. Assume $U(\partial_\mu \varphi\partial^\mu \varphi^*)$
can be written as:

\begin{equation}\label{49}
U(\partial_\mu \varphi\partial^\mu \varphi^*)=\lambda
[\partial_\mu \varphi\partial^\mu
\varphi^*-\frac{1}{V_\Omega}\int\limits_\Omega
\partial_\mu \varphi\partial^\mu \varphi^* \mathrm{d}V].
\end{equation}                        %(49)

where $\lambda$ is called the coupling coefficient and $V_\Omega$
is the volume of $\Omega$. It is easy to show that

\begin{equation}\label{50}
\int\limits_\Omega U(\partial_\mu \varphi\partial^\mu \varphi^*)
\mathrm{d}V=0.
\end{equation}                        %(50)

Due to the equality above, $U(\partial_\mu \varphi\partial^\mu
\varphi^*)$ may be interpreted as a fluctuation of self-energy of
field over the space-time domain $\Omega$. Clearly, both the
Lagrangian and its action integral are invariant under the global
gauge transformation

\begin{equation}\label{51}
\varphi \to e^{-i\phi}\varphi,\mspace{16mu} \varphi^* \to
e^{i\phi}\varphi^* ,
\end{equation}                         %(51)

where $\phi$ is a real constant. For a long time, there are two ways
of comprehending the gauge invariance of a Lagrange system
\cite{s5}. One is based on the invariance of the Lagrangian; the
other is based on the invariance of the action integral of the
Lagrangian. When the localized assumption is not valid, the gauge
invariance of the Lagrangian is included in the gauge invariance of
the action integral. So under the gauge transformation, it is of
more generality to comprehend the invariance of a Lagrange system as
the invariance of the action integral. Let

\begin{equation}\label{52}
L_0=\partial_\mu \varphi \partial^\mu
\varphi^*-m^2\varphi\varphi^*-V(\varphi\varphi^*).
\end{equation}                        %(52)

Then Eq.(\ref{48}) can be written as

\begin{equation}\label{53}
L=L_0-U(\partial_\mu \varphi \partial^\mu \varphi^*).
\end{equation}                        %(53)

Due to Eq.(\ref{50}), {\itshape{L}} and $L_0$ have the same action
integral on $\Omega$. Therefore, the Euler-Lagrange equations
derived from {\itshape{L}} and $L_0$ have the same expression,
which read

\begin{equation}\label{54}
(\Box+m^2)\varphi=-\frac{\partial V}{\partial\varphi^*},
\mspace{16mu}(\Box+m^2)\varphi^*=-\frac{\partial
V}{\partial\varphi},
\end{equation}                       %(54)

which are two Klein-Gordon equations. The infinitesimal form of
the transformations (\ref{51}) can be represented as

\begin{equation}\label{55}
\delta\varphi =-i\phi\varphi,\mspace{16mu} \delta\varphi^* =
i\phi\varphi^* ,
\end{equation}                         %(55)

and so

\begin{equation}\label{56}
\delta(\partial^\mu\varphi)
=-i\phi(\partial^\mu\varphi),\mspace{16mu}
\delta(\partial^\mu\varphi^*) = i\phi(\partial^\mu\varphi^* ).
\end{equation}                          %(56)

Under these infinitesimal transformations, as the same as deriving
Eq.(\ref{54}), we can obtain the equality below:

\begin{equation}\label{57}
\mathit{\int\limits_\Omega i[\varphi^* \frac{\partial L}{\partial
(\partial_\mu\varphi^*)} - \varphi \frac{\partial L}{\partial
(\partial_\mu\varphi)}]_{,\mu}\mathrm{d}V=\int\limits_\Omega
i[\varphi^* \frac{\partial L_0}{\partial (\partial_\mu\varphi^*)}
- \varphi \frac{\partial L_0}{\partial
(\partial_\mu\varphi)}]_{,\mu}\mathrm{d}V}.
\end{equation}                       %(57)

With Eq.(\ref{52}), the right-hand term of Eq.(\ref{57}) is
written as

\begin{equation}\label{58}
\mathit{\int\limits_\Omega i[\varphi^* \frac{\partial
L_0}{\partial (\partial_\mu\varphi^*)} - \varphi \frac{\partial
L_0}{\partial
(\partial_\mu\varphi)}]_{,\mu}\mathrm{d}V=\int\limits_\Omega
i(\varphi^* \partial_\mu \partial^\mu\varphi - \varphi
\partial_\mu\partial^\mu\varphi^*)\mathrm{d}V}.
\end{equation}                       %(58)

It follows immediately from Eq.(\ref{54}) that we have

\begin{equation}\label{59}
\mathit{\int\limits_\Omega i[\varphi^* \frac{\partial
L_0}{\partial (\partial_\mu\varphi^*)} - \varphi \frac{\partial
L_0}{\partial
(\partial_\mu\varphi)}]_{,\mu}\mathrm{d}V=\int\limits_\Omega
i(\varphi \frac{\partial V}{\partial_\mu\varphi} - \varphi^*
\frac{\partial V}{\partial_\mu\varphi^*})\mathrm{d}V}=0.
\end{equation}                       %(59)

Let

\begin{equation}\label{60}
\mathit{J^\mu= i[\varphi^* \frac{\partial L}{\partial
(\partial_\mu\varphi^*)} - \varphi \frac{\partial L}{\partial
(\partial_\mu\varphi)}]} ,
\end{equation}                       %(60)

which denotes the conservation flux of {\itshape{L}}. Thus,
Eq.(\ref{57}) becomes

\begin{equation}\label{61}
\mathit{\int\limits_\Omega
J_{,\mu}^\mu\mathrm{d}V=\int\limits_\Omega i[\varphi^*
\frac{\partial L_0}{\partial (\partial_\mu\varphi^*)} - \varphi
\frac{\partial L_0}{\partial
(\partial_\mu\varphi)}]_{,\mu}\mathrm{d}V}.
\end{equation}                       %(61)

Inserting Eq.(\ref{59}) in (\ref{61}) yields

\begin{equation}\label{62}
\mathit{\int\limits_\Omega J_{,\mu}^\mu\mathrm{d}V}=0.
\end{equation}                       %(62)

If the localized assumption holds, from Eq.(\ref{62}) we
immediately obtain

\begin{equation}\label{63}
\mathit{J_{,\mu}^\mu= i[\varphi^* \frac{\partial L}{\partial
(\partial_\mu\varphi^*)} - \varphi \frac{\partial L}{\partial
(\partial_\mu\varphi)}]_{,\mu}}=0.
\end{equation}                       %(63)

However, substituting Eq.(\ref{48}) into (\ref{63}), we have

\begin{align}\label{64}
&\mathit{J_{,\mu}^\mu=i[\varphi^*\frac{\partial L}{\partial
(\partial_\mu\varphi^*)}-\varphi\frac{\partial L}{\partial(\partial_\mu\varphi)}]_{,\mu}}\notag\\
&=\mathit{\frac{i
\lambda}{V_\Omega}[\varphi^*\frac{\delta}{\delta(\partial_\mu\varphi^*)}\int\limits_\Omega
\partial_\nu\varphi^*\partial^\nu\varphi\mathrm{d}V-\varphi
\frac{\delta}{\delta(\partial_\mu\varphi)}\int\limits_\Omega
\partial_\nu\varphi\partial^\nu\varphi^*\mathrm{d}V]_{,\mu}}\notag\\
&=\mathit{\frac{i\lambda}{V_\Omega}(\partial_\mu\varphi^*
\int\limits_\Omega\partial^\mu\varphi\mathrm{d}V-\partial_\mu\varphi
\int\limits_\Omega\partial^\mu\varphi^*\mathrm{d}V)},
\end{align}                      %(64)

where $\delta / \delta (\cdot)$ refers to the Frechet derivative.
In general, the right-hand term of Eq.(\ref{64}) is not equal to
zero. Therefore, Eq.(\ref{63}) is contradicted with Eq.(\ref{64}).
---This only shows the localized assumption is on longer valid.
Consequently, By introducing the nonlocal residual, Eq.(\ref{62})
can be transformed to the differential equation below:

\begin{equation}\label{65}
\mathit{J_{,\mu}^\mu= F}.
\end{equation}                       %(65)

Comparing Eq.(\ref{64}) with (\ref{65}) gives rise to

\begin{equation}\label{66}
\mathit{F = \frac{i \lambda}{V_\Omega}(\partial_\mu\varphi^*
\int\limits_\Omega\partial^\mu\varphi\mathrm{d}V-\partial_\mu\varphi
\int\limits_\Omega \partial^\mu\varphi^* \mathrm{d}V)}.
\end{equation}                       %(66)

As shown in Eq.(\ref{66}), {\itshape{F}} has a anti-symmetry with
respect to $\partial^\mu\varphi$ and $\partial_\mu\varphi^*$. So
we can easily verify that it satisfies the zero mean condition.
That is

\begin{equation}\label{67}
\mathit{\int\limits_\Omega F(x^\mu ) \mathrm{d}V}=0.
\end{equation}                       %(67)

With Eq.(\ref{45}), the nonlocal residual is also represented as

\begin{equation}\label{68}
\mathit{F= ie(\mathrm{1}+\frac{\delta
U}{\delta\theta})[(\partial^\mu\varphi)A_\mu\varphi^*-(\partial^\mu\varphi^*)A_\mu\varphi]},
\end{equation}                       %(68)

where $\theta=\partial_\mu \varphi\partial^\mu \varphi^*$, being an
intermediate variable. Inserting Eq.(\ref{68}) in (\ref{66}) leads
to

\begin{equation}\label{69}
\mathit{e(\mathrm{1}+\frac{\delta
U}{\delta\theta})[(\partial^\mu\varphi)A_\mu\varphi^*-(\partial^\mu\varphi^*)A_\mu\varphi]
=\frac{\lambda}{V_\Omega}(\partial_\mu\varphi^*
\int\limits_\Omega\partial^\mu\varphi\mathrm{d}V-\partial_\mu\varphi
\int\limits_\Omega \partial^\mu\varphi^* \mathrm{d}V)},
\end{equation}                       %(69)

from which we immediately obtain

\begin{equation}\label{70}
\mathit{A_\mu\varphi=-\frac{\lambda}{eV_\Omega}(\mathrm{1}+\frac{\delta
U}{\delta\theta})^{\mathrm{-1}}\int\limits_\Omega\partial_\mu\varphi\mathrm{d}V,
\mspace{8mu}
A_\mu\varphi^*=-\frac{\lambda}{eV_\Omega}(\mathrm{1}+\frac{\delta
U}{\delta\theta})^{\mathrm{-1}}\int\limits_\Omega\partial_\mu\varphi^*\mathrm{d}V}.
\end{equation}                       %(70)

Eq.(\ref{70}) characterizes correlations between the gauge field
$A_\mu$ and the gradient of the Lagrangian field
$\partial_\mu\varphi$ under the case of global gauge invariance
concerned with nonlocal effects. Obviously, if
$U=U(\partial_\mu\varphi\partial^\mu\varphi^*)=0$, then we easily
verify $J_{,\mu}^\mu=0$ from Eq.(\ref{48}), (\ref{54}) and
(\ref{60}). This shows that no nonlocal effect will exist,
provided no fluctuation of the self-energy of field occurs. So
far, we have seen that although the fluctuation of the self-energy
of field has no influence on the motion equation and the symmetry
of the action integral defined on the global domain, it enables to
locally break the conservation flux so that it could not remain
constant.
---This is an observable effect. We expect that the experiment in
the future can confirm existence of this effect.

\section{Conclusions}
Under the condition that the localized assumption is no longer
valid, the connection between the conservation flux and the gauge
invariance of the action integral under the finite Lie group of
infinitesimal transformations is established. This is the so-called
nonlocal balance equation. It shows that divergence of the
conservation flux is equal to the nonlocal residual rather than
zero. ---Due to this fact, the nonlocal balance equation is not a
conservation law in a strict sense, but it should be regarded as a
generalization of the Noether's theorem under the global gauge transformation. \\
The nonlocal residual is subjected to the constraint of the zero
mean condition. Therefore, it has no influences on the motion
equation and the gauge invariance of the action integral defined on
a bounded domain $\Omega$ as a whole. However, the nonlocal residual
enables to locally break the conservation flux so that it no longer
remains constant in the local sub-domain of $\Omega$. This is an
observable effect. In our opinion, some special experiments should
be able to certify it. \\
Physically, the nonlocal residual may be interpreted as a new
source coming from interactions within a matter or field. ---On
the basis of this argument, the correlation between the nonlocal
residual and the gauge field is naturally established, just as
shown by Eq.(\ref{45}). This result also shows that when the
localized assumption is no longer valid, if and only if the gauge
field is introduced, the global gauge invariance of a Lagrangian
system can be accurately
characterized. \\
The example given in this paper further verifies that the localized
assumption is no avail under some circumstance. If such situations
occur, the global gauge invariance of the action integral can be
described by the nonlocal balance equation, by which the nonlocal
residual and the gauge field are also determined explicitly.
Meanwhile, this example also illustrates that the localized
assumption probably fails only when self-interactions of field
occur. For a free field, the localized assumption is always valid.

\section*{Acknowledgements}
The support of the National Nature Science Foundation of China
through the Grant No. 10472135 is gratefully acknowledged.

\end{document}